\documentclass[aps,prd,preprint,groupedaddress,showkeys]{revtex4-1}

 \usepackage{graphicx} 
\usepackage{amsmath}
\usepackage{amsfonts}

\newcommand{\bo}{\raise-1mm\hbox{\Large$\Box$}}

\begin{document}

\title{Probing AdS/QCD backgrounds with  semi-classical strings}



\author{Saulo Diles}
\email{smdiles@ufpa.br}
\affiliation{Campus Salinópolis, Universidade Federal do Par\'{a},
68721-000, Salin\'{o}polis, Par\'{a}, Brazil}

\begin{abstract}

 New AdS/QCD backgrounds have been proposed to describe the spectrum of heavy vector mesons via 
 the implementation of additional energy scales on the bulk geometry of the soft wall model. The extra energy scales are needed to include the decay constants of hadronic states when describing the radial excitations of the heavy meson. Here we analyze one model that  introduces an ultraviolet cutoff on Anti de-Sitter space and a  model  that consider a  dilaton profile modified by the addition of an extra term and no cutoff.
For each one of these two models we consider the presence of a  semi-classical string in the bulk that is dual to  a static and infinitely heavy meson. We compute the expected value of the Wilson loop operator using the holographic dictionary and obtain the dual potential for the static $q\bar{q}$ pair.  For the model with modified dilaton profile the on-shell string action presents a peculiar ultraviolet divergence, a compatible regularization is  discussed and a new subtraction scheme is used. We consider the case of finite temperature and determine how the dissociation temperature of the heavy meson is affected by the additional energy scales.  

\end{abstract}


\maketitle

\section{ Introduction }  
 The  AdS/CFT correspondence \cite{Maldacena:1997re,Witten:1998qj,Rey:1998ik,Maldacena:1998im, Nastase:2007kj,Ramallo:2013bua}  allows for treating the strong coupling regime of a non-abelian gauge 
 theory.  With the rise of Lattice QCD   \cite{Wilson:1974sk,Kogut:1974ag, Kogut:1982ds}  the phenomenology of non-abelian gauge theories  have been explored more directly. 
  Lattice simulations shows that, in a confining gauge theory, the interaction potential is linear 
  for large distances,  $U(r)\sim \sigma r + \mathcal{O}(1)$, while for any CFT the conformal symmetry requires that the  potential  be   Coulomb type, 
   $U_{_\textit{CFT}}(r)\sim \frac{1}{r},$ and
 there is no confinement.   A first proposal that modifies the AdS/CFT dictionary  to describe QCD phenomenology was made by Polchinski and Strassler in \cite{Polchinski:2002jw}, where the
  conformal symmetry is broken through the
 introduction of an infrared wall and it leads to a good description of the cross sections in deep inelastic scattering. The same strategy was used to describe glueball
 mass spectrum \cite{BoschiFilho:2002ta,BoschiFilho:2002vd}.
 This scenario, known as the Hard-Wall model, was used to compute Wilson loops of static mesons   \cite{BoschiFilho:2004ci,Boschi-Filho:2006ssg, BoschiFilho:2006pe}. 

 It was proposed in \cite{Karch:2006pv} a different way to introduce an energy scale in the model by coupling  bulk fields with a background dilaton,
establishing the soft wall AdS/QCD model. This AdS/QCD background was used to describe static mesons using open strings in the bulk  \cite{Andreev:2006ct, Andreev:2006eh}. 
The spectral functions of the $\rho$ meson were computed using the soft wall model in \cite{Fujita:2009wc}.
Soft wall model is successful in describing the radial excitations of light vector mesons, 
but the same does not happen for the radial excitations of the heavy charmonium and bottomonium  mesons. 
The current experimental data \cite{Tanabashi:2018oca} provides the decay constants for these   states observing they are decreasing with the excitation number.
However, the theoretical prediction of the soft wall model is that the decay constants are the same for all radial excitations. 
A first tentative to describe the observed decay constants using  AdS/QCD  was done in \cite{Grigoryan:2010pj}.
This reference \cite{Grigoryan:2010pj} motivated the search for a description of decay constants using  AdS/QCD and called attention for the role of decay constants in the behavior of the heavy mesons at finite temperature. 
It was constructed  an AdS/QCD model where an ultraviolet cutoff is introduced in  the bulk geometry of the soft wall model  \cite{Evans:2006ea,Afonin:2011ff, Braga:2015jca}. The correlation functions  at  zero temperature  obtained in this AdS/QCD model is  consistent with the dependence of decay constants of charmonium and  bottomonium with the radial excitation level.
Effects of finite temperature were considered in \cite{Braga:2016wkm}, the obtained spectral functions provide a good description of 
the dissociation of bottomonium states while for the charmonium states  the peaks of the spectral function disappear at low temperatures and the dissociation temperature of 
the 1S state ($J/\Psi$) is underestimated by the model.  This AdS/QCD background with ultraviolet cutoff was also used to calculate spectral functions  
in the case of finite  density  in   \cite{Braga:2017oqw, Braga:2017wpy} and  to describe the mass spectrum of light mesons \cite{Cortes:2017lgz}. 

It was proposed in \cite{Braga:2017bml} an AdS/QCD model where the  dilaton is modified by  the inclusion of two parameters. 
In this model there is no hard cutoff and it leads to  very good  decay constants for the first four  radial  excitations of  charmonium. 
The obtained spectral functions also show a consistent description of the dissociation for the charmonium states. The model proposed in \cite{Braga:2017bml} was refined and a more complete AdS/QCD model for heavy vector mesons was reached in Ref. \cite{Braga:2018zlu}. The AdS/QCD background of \cite{Braga:2018zlu} is very precise in describing the decay constants of charmonium and bottomonium.

In the present paper these AdS/QCD backgrounds (\cite{Evans:2006ea,Afonin:2011ff, Braga:2015jca} and \cite{Braga:2017bml, Braga:2018zlu}) are probed by semi-classical static strings. 
The string is dual to a static meson of the  gauge theory and the holographic dictionary gives the expectation value of the Wilson loop operator.  Both models present an infrared wall in the AdS that is responsible 
for confinement at zero temperature. 
 
The paper is organized as follows: in section II  we analyze the string in the AdS/QCD background  with an ultraviolet cutoff, 
obtain the $q\bar{q}$ potential and explore the dissociation of the meson at finite temperature. 
In section III we analyze the string in the AdS/QCD background with a modified dilaton profile proposed in references  \cite{Braga:2017bml,Braga:2018zlu},  discuss the regularization of the string action proposing an alternative scheme that we use to obtain the $q\bar{q}$ potential  and explore the relation between the dissociation of the meson with the additional energy scales of the models. Section IV is dedicated to a discussion of the obtained results.

\section{Soft-wall with  ultraviolet cutoff}
\begin{figure}[h] 
	\centering
		\includegraphics[width=0.45 \textwidth]{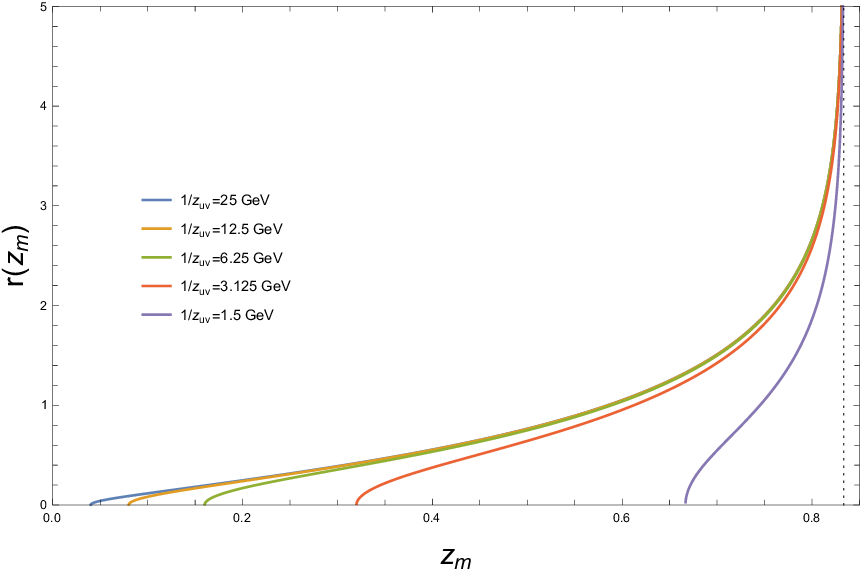}
	\caption{Separation distance as a function of $z_m$.}
	\label{fig:fig1}
\end{figure}

In this section we analyze the AdS/QCD background of Refs. \cite{Evans:2006ea,Afonin:2011ff, Braga:2015jca}, where 
two point correlation functions are evaluated on the slice of AdS at finite $z=z_{uv}$.
We consider that a static string  in this background, with its endpoints attached on the slice at $z_{uv}$, is dual to the static 
quark/ anti-quark pair. We use Poincar\'{e} coordinates $(t,\vec{x},z)$ to describe the Euclidean geometry of the deformed AdS geometry in the string frame. The metric is
\begin{equation}
 ds^2 = R^2\frac{e^{\phi(z)}}{z^2}(dt^2 + d\vec{x}^2 + dz^2).\label{metric1}
\end{equation}
The holographic coordinate is $z\in [z_{uv}, \infty)$ and the dilaton profile is quadratic in the holographic direction, i.e., $\phi(z)=k^2z^2.$ 
The  action of the string is the Nambu-Goto action
\begin{equation}
 S = \frac{1}{2\pi \alpha'}\int d^2\sigma \sqrt{|det(g_{ij}\partial_\alpha X^i\partial_\beta X^j)|}, 
\end{equation}
where $g_{ij}$ is the metric in eq.(\ref{metric1}). The string endpoints are attached to the locations of the quark and the anti-quark
at $\vec{x}_q=(\frac{r}{2},0,0)$ and $\vec{x}_{\bar{q}}=(-\frac{r}{2},0,0).$  
 The symmetries of the bulk geometry provide a simple parameterization of the world-sheet, namely  $X^i = (t, x,0,0, z(x)),~\sigma^0=t,~\sigma^1=x$.
We set  from now the coupling $g=\frac{R^2}{\alpha'}=1$. Therefore, the string action is 
 \begin{equation}
 S = \frac{1}{2\pi }\int_0^T dt \int_{-\frac{r}{2}}^{\frac{r}{2}} dx \frac{e^{\phi(z)}}{z^2}\sqrt{1+z'^2}.
 \end{equation} 
  The world-sheet configuration is found using the classical mechanics approach described in
\cite{Kinar:1998vq}. We assume that the function $z(x)$ reach its maximum at $z_m$ and we use the fact that the Hamiltonian 
\begin{equation}
 \mathcal{H} = z'\frac{\partial \mathcal{L}}{\partial z'} -\mathcal{L} = \frac{ e^{k^2z^2}}{z^2\sqrt{1+z'^2}} = \frac{e^{k^2z_m^2}}{z_m^2}
\end{equation}
is a constant of motion. It happens that in this background the string endpoints do not reach the boundary at $z_{uv}$ orthogonally  due to the conservation of this Hamiltonian.
  Here we are not interested in the details of the world-sheet but on its  property that for a given separation of its endpoints the string 
 reaches the maximum position $z_m$ on the bulk. The distance between the quark and anti-quark is then obtained as a function of $z_m$: 
 \begin{equation}
  r(z_m) = 2 z_m\int_{\frac{z_{uv}}{z_m}}^1 dv \frac{v^2e^{k^2 z_m^2(1-v^2)}}{\sqrt{1-v^4e^{2k^2z_m^2(1-v^2)}}}. \label{sepdistance1}
  \end{equation}  
 In Figure \ref{fig:fig1} we fix $k=1.2~ GeV$ and we show the behavior of $r(z_m)$ for five different locations of the ultraviolet cutoff, in all plots the distances are in units of GeV$^{-1}$.
 The infrared wall is represented in Figure \ref{fig:fig1} by the dotted vertical line at $z_{uv}=1/k$, the global minimum of $f(z) = e^{2k^2z^2}/z^4$. For each value of the dilaton coupling $k$, the combination  $k^2z_{uv}^2$ cannot reach the unit so  $z_{uv} < \frac{1}{k}.$  
 This separation of scales in the model was speculated in Ref. \cite{Braga:2015jca} and is confirmed by the present analysis.

The dual  potential for the $q\bar{q}$ pair is expressed as a function of $z_m$  by
\begin{equation}
 U(z_m) = \frac{1}{\pi z_m}\bigg\{-1 + \int_{\frac{z_{uv}}{z_m}}^1 dv\frac{e^{\phi(vz_m)}}{v^2}\left(\frac{1}{\sqrt{1-v^4e^{2[\phi(z_m)-\phi(v z_m)]} }} -1\right) \bigg\}. \label{regpotential1}
\end{equation}
The regularization scheme used here is not usual and its details are discussed in the next section. 

\begin{figure}[h] 
 		\includegraphics[width=0.45 \textwidth]{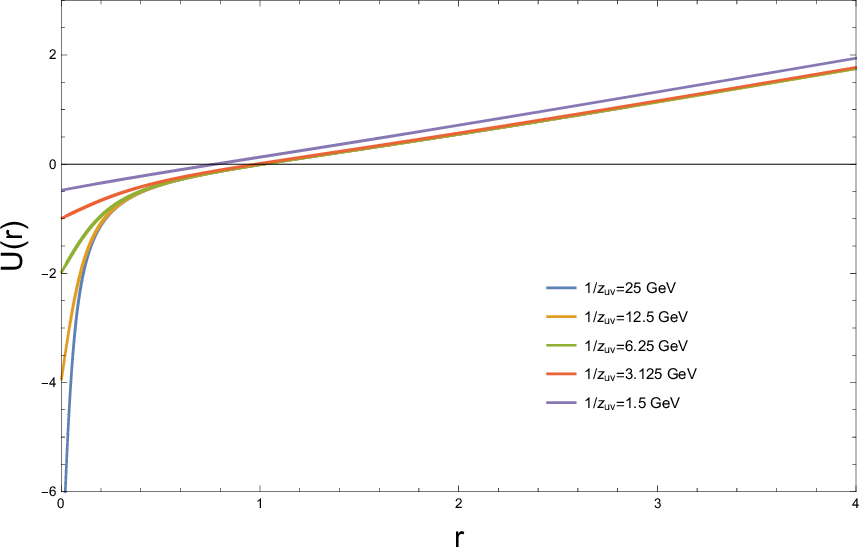}
	\caption{Potential $U(r)$ for different $z_{uv}$.}
	\label{fig:fig2}
\end{figure}

We use  eq.(\ref{sepdistance1}) and eq.(\ref{regpotential1}) to perform a parametric plot of the potential as a function of $q\bar{q}$ separation.  We show in Figure \ref{fig:fig2} the curves $U(r)$, where we fix $k=1.2 GeV$ and take five representative values of the ultraviolet cutoff, in all plots the energies are in units of GeV.  
The effect of the cutoff for the dual potential is dramatic: the  potential is no longer Coulomb-like in the $r\to0$ limit, instead it reaches a  finite negative value. The ultraviolet cutoff in the AdS space translates into a finite $q\bar{q}$ potential potential in the origin.
It is important to remark that as we take smaller values for $z_{uv}$   the constant value of the regularized potential at $r=0$ assumes greater negative values. The 
Coulomb potential in the ultraviolet region is recovered in the limit $z_{uv}\to 0$. This limit  corresponds precisely to the soft wall model. It is also notable that the string tension is low sensitive with the ultraviolet cutoff.

\subsection{Finite temperature}
The gauge theory at finite temperature  is dual to a black hole geometry in the five dimensional Anti-de Sitter space. The Euclidean black hole geometry of the present model reads  

\begin{equation}
 ds^2 = R^2\frac{e^{\phi(z)}}{z^2}\left(f(z)dt^2 + d\vec{x}^2 + \frac{1}{f(z)}dz^2\right),
\end{equation}

\noindent where $f(z) = 1-\frac{z^4}{z_h^4}$ and there is an event horizon on $z_h,$ so $z\in[z_{uv},z_h)$. The temperature of the dual gauge theory is given by
\begin{equation}
 T = \frac{1}{\pi z_h\sqrt{f(z_{uv})}}, \label{tdual}
\end{equation}
as explained in Ref. \cite{Braga:2016wkm}.
The dissociation temperature of the heavy meson  is obtained using the procedure described in ref.  \cite{Colangelo:2010pe}, that consists in  analyzing  the behavior of the function $r(z_m)$ when decreasing the location of the event horizon. At the critical value,  $z_h=z_{critical}$, the curve $r(z_m)$ ceases to be disconnected to become connected 
denouncing the phase transition. The dissociation temperature is obtained setting $z_h=z_{critical}$ in  eq.(\ref{tdual}).  For the model analyzed in this section we have   
\begin{equation}
r(z_m) = 2 z_m\int_{\frac{z_{uv}}{z_m}}^1 dv \frac{v^2e^{k^2 z_m^2(1-v^2)}}{f(vz_m)\sqrt{1-v^4e^{2k^2z_m^2(1-v^2)}\frac{f(z_m)}{f(vz_m)}}}. 
\end{equation}
It comes out that  $z_{critical}$ does not coincide with the location of the infrared wall. Instead we found that  $z_{critical}> z_{wall}$.  
In this case, there is no analytic expression for the dissociation temperature  as a function of the model parameters.
The dependence of the dissociation temperature on the cutoff scale $z_{uv}$ is explored for a set of representative values of $z_{uv}$.
\begin{figure}[h]
	\centering
		\includegraphics[width=0.45 \textwidth]{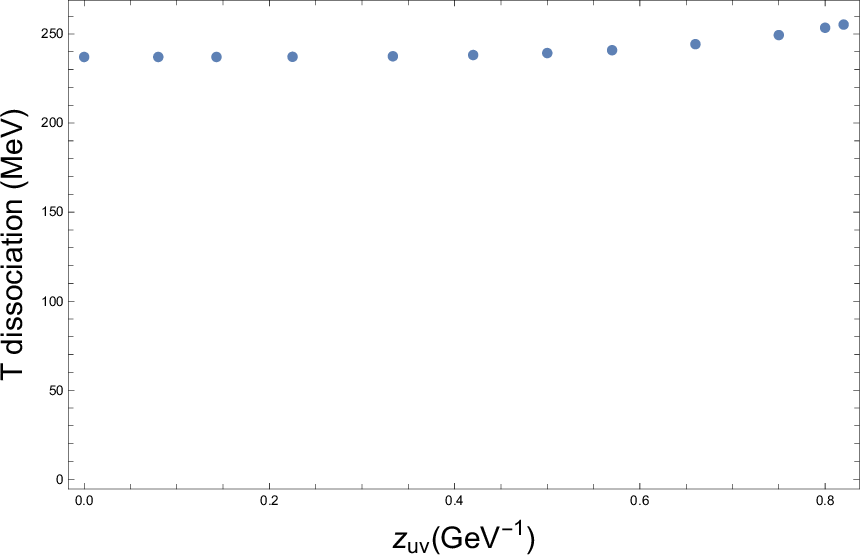}
	\caption{Dissociation temperature as a function of $z_{uv}$.}
	\label{fig:fig3}
\end{figure}
We show the results in Figure \ref{fig:fig3}. The dissociation temperature is low sensitive with the location of the ultraviolet cutoff.

 \section{The modified dilaton profile}

It was proposed in \cite{Braga:2017bml}  an AdS/QCD model where the quadratic dilaton profile is deformed by 
the addition of an extra term. In the model of Ref. \cite{Braga:2017bml} the dilaton profile is
\begin{equation}
 \Phi(z) = k^2z^2+ tanh \left( \frac{1}{Mz}-\frac{k}{\sqrt{\Gamma}} \right).  \label{dilaton} 
\end{equation}
In this  model the observed  spectrum of masses and  decay constants of  charmonium is  fitted by taking the model parameters as $k=1.2$GeV, $M= 2.7$GeV, $\sqrt{\Gamma}=0.75$GeV, that are used here as reference values. W analyze the AdS/QCD background of this model by  considering a static string in the bulk. For the present case, the Euclidean geometry in the string frame is 
\begin{equation}
 ds^2 = R^2\frac{e^{\Phi(z)}}{z^2}(dt^2 + d\vec{x}^2 + dz^2),~~z\in (0,\infty).\label{metric}
\end{equation}
The dual gauge theory lives in the conformal boundary at $z= 0$.
Classical equations of motion for the world-sheet gives the $q\bar{q}$  separation as a function $z_m$:
\begin{equation}
 r(z_m) = 2 z_m \int_0^1 dv \frac{v^2}{\sqrt{e^{2[\Phi(vz_m) - \Phi(z_m)]}-v^4}}. 
\end{equation} 
\begin{figure}[h] 
	\includegraphics[width=0.45 \textwidth]{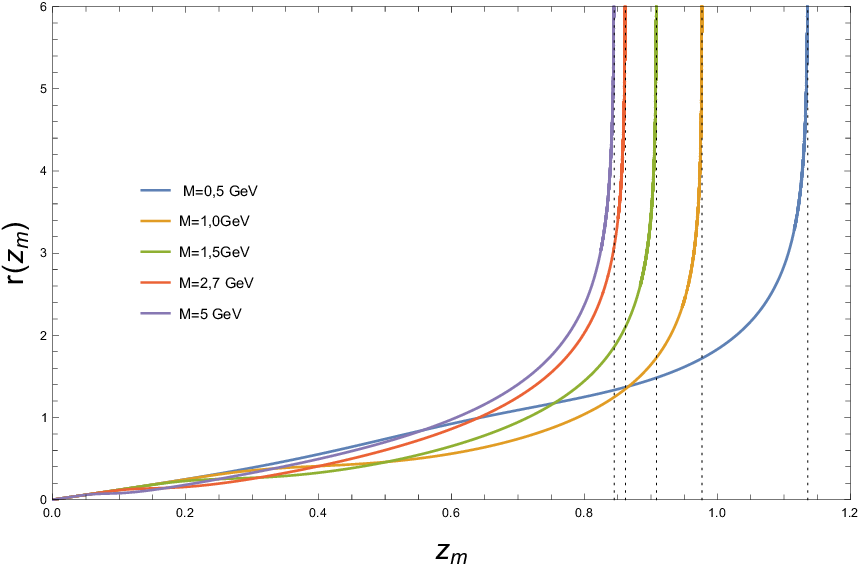}
    \caption{The infrared wall as a function of $M.$}
    \label{fig:fig4.1}
  \end{figure}
  \begin{figure}[h]
	\includegraphics[width=0.45 \textwidth]{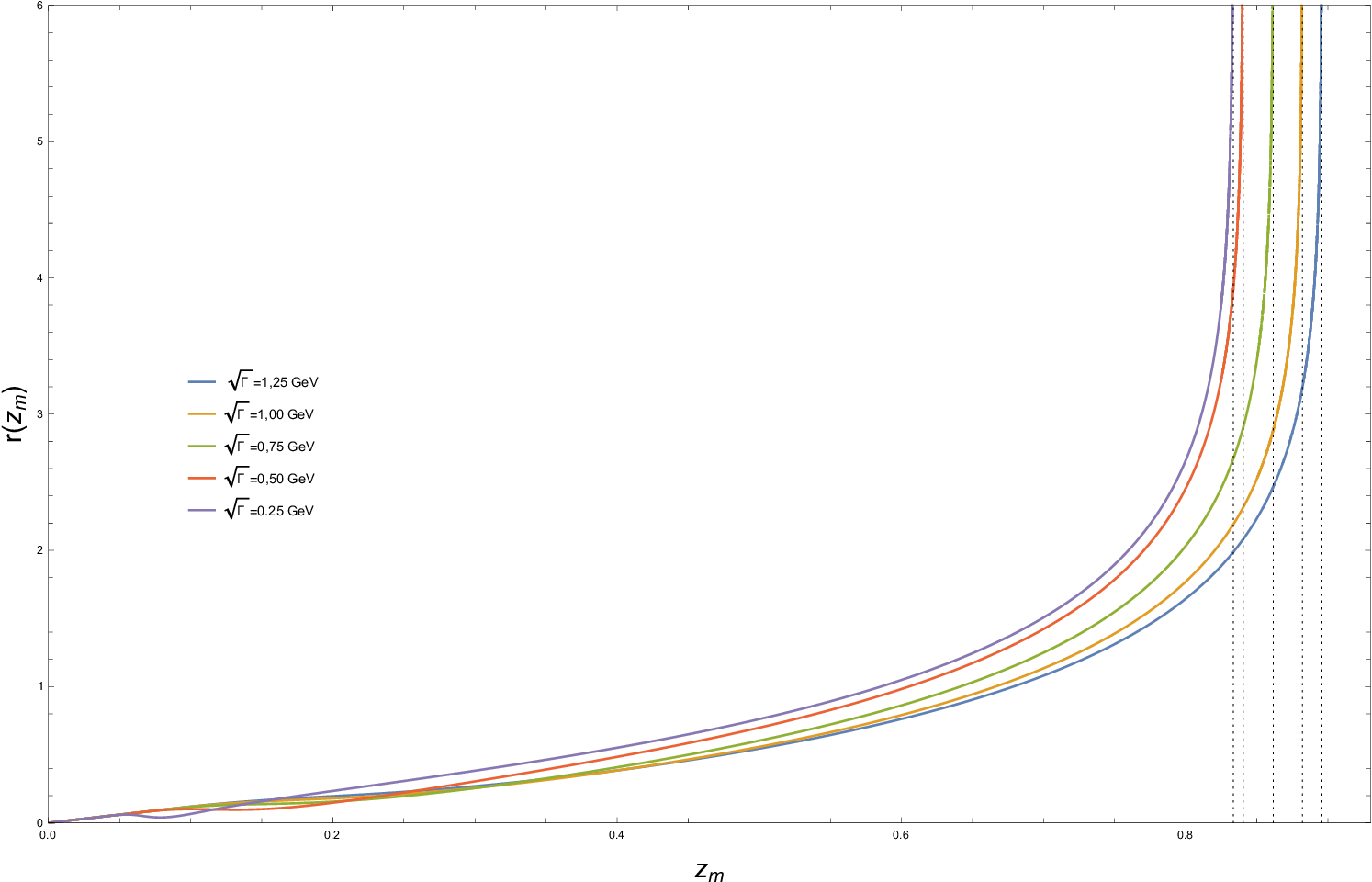}
    \caption{The infrared wall as a function of $\sqrt{\Gamma}.$ }
    \label{fig:fig4.2}
  \end{figure}

We show the $q\bar{q}$ separation as a function of  $z_m$ for a set of different model parameters
in Figure \ref{fig:fig4.1}, where we vary M and in Figure \ref{fig:fig4.2}  varying $\sqrt{\Gamma}$. The location of the infrared wall is determined by the local minumum of $f_\Phi(z) = e^{2\Phi(z)}/z^4$ and is represented in  Figure \ref{fig:fig4.1} and \ref{fig:fig4.2} by dotted vertical lines.
In this model the location of the infrared wall is sensitive on the extra energy scales. 
This sensitivity on the shape of the world-sheet on the additional energy scales $M$ and $\sqrt{\Gamma}$ of the model analyzed in this section is similar to what is found when analyzing  the world-sheet of a rotating string
in AdS geometry when corrections due to a finite t'Hooft coupling and a Gauss-Bonnet term are considered, see Ref. \cite{AliAkbari:2009pf}.


The computation of the $q\bar{q}$ potential requires a  regularization of the divergent integral of the classical string action. In the original proposal \cite{Rey:1998ik, Maldacena:1998im} 
the regularization is done by subtracting twice the action of a string stretched from $z=0$ up to $z\to\infty$ with $(t,\vec{x})$ coordinates keep constant. This procedure is interpreted as 
subtracting the divergent masses of the infinitely heavy quarks and is generalized for arbitrary geometries in \cite{Kinar:1998vq}. The regularization of UV divergences that appear in Wilson Loop integral have been extensively explored in the context of AdS/CFT, see Refs.  \cite{Drukker:1999zq,Drukker:2005kx,Chu:2008xg}. 
In the soft wall model, the divergent action integral is regularized in  \cite{Andreev:2006ct} subtracting $\frac{1}{2\pi}\int_0^\infty \frac{dz}{z^2}$, that ignores the presence of the dilaton. This regularization cannot be interpreted as subtracting the divergent masses of the constituent quarks in the sense of \cite{Kinar:1998vq}. In the soft wall model, the action for a string representing the single quark  is $\frac{1}{\pi}\int_0^\infty dz\frac{ e^{k^2z^2}}{z^2}$, that is  divergent for $z\to \infty$ due to the presence of the dilaton. Despite the interpretation of quark mass, the integrals with and without the dilaton diverges in the same way in the region $z\to 0$, since the dilaton vanishes there, and the ultraviolet divergence is precisely canceled. 
For the background we analyze in this section, it happens that the dilaton does not vanishes on the boundary, rather we face that  $\Phi(z\to0)\to 1$. Because of this the regularization used in \cite{Andreev:2006ct} cannot be adopted, in order to precisely cancel the ultraviolet divergence in the string action one cannot ignore the presence of the dilaton near $z\to 0$. The regularization defined in  Ref. \cite{Kinar:1998vq} also does not apply here since the integral prescribed  to be subtracted, $\frac{1}{\pi}\int_0^\infty dz\frac{ e^{\Phi(z)}}{z^2}$,  is divergent for $z\to \infty$ while the action of the string representing the $q\bar{q}$ pair avoids this divergence due to the infrared wall at $z_m$. So, if we use the prescription of Ref. \cite{Kinar:1998vq} the ultraviolet divergence is cancelled but an infrared one is generated. The regularization procedure must cancel precisely the ultraviolet divergence without generating an infrared one. Here we propose one alternative procedure to
regularize the quark/anti-quark potential by subtracting  the divergent expression 
\begin{equation}
\frac{1}{\pi}\int_{0}^{z_m} dz \frac{e^{\Phi(z)}}{z^2}+\frac{1}{\pi}\int_{z_m}^\infty dz  \frac{1}{z^2}, \label{divmass}
\end{equation} 
that in the ultraviolet  diverges in the same way as the world-sheet action and is finite in the infrared. Using this regularization scheme we obtain the dual potential as a function of $z_m$ 
\begin{equation}
 V(z_m) = \frac{1}{\pi z_m}\bigg\{-1 + \int_0^1 dv\frac{e^{\Phi(vz_m)}}{v^2}\left(\frac{1}{\sqrt{1-v^4e^{2[\Phi(z_m)-\Phi(v z_m)]} }} -1\right) \bigg\}.
\end{equation}
 The regularization of the Wilson Loop integral in non-conformal holography was previously discussed in Refs. \cite{Quevedo:2013iya, Pontello:2015yla} where a slightly different scheme was proposed. Their strategy is to place an infrared cutoff at $z=z_m$ in such a way that the term they subtract corresponds precisely to the first integral in eq.(\ref{divmass}). If one use the regularization scheme of  \cite{Quevedo:2013iya, Pontello:2015yla} in the soft-wall model, the obtained $q\bar{q}$ potential will not be Coulomb-like for $r\to0$, rather it will present a positive divergence there. The second term in eq.(\ref{divmass})  fix this issue and when the present regularization is applied to the soft wall model one obtain  a $q\bar{q}$ potential of the Cornell type.
 The present proposal agree with the idea presented in Refs. \cite{Quevedo:2013iya, Pontello:2015yla} in using $z_m$ as a natural scale to establish the regularization.  We also point out that the subtraction scheme discussed here was implicitly adopted in Ref. \cite{Bruni:2018dqm}, where the Cornell potential was obtained.  

  \begin{figure}[h]
	\includegraphics[width=0.45 \textwidth]{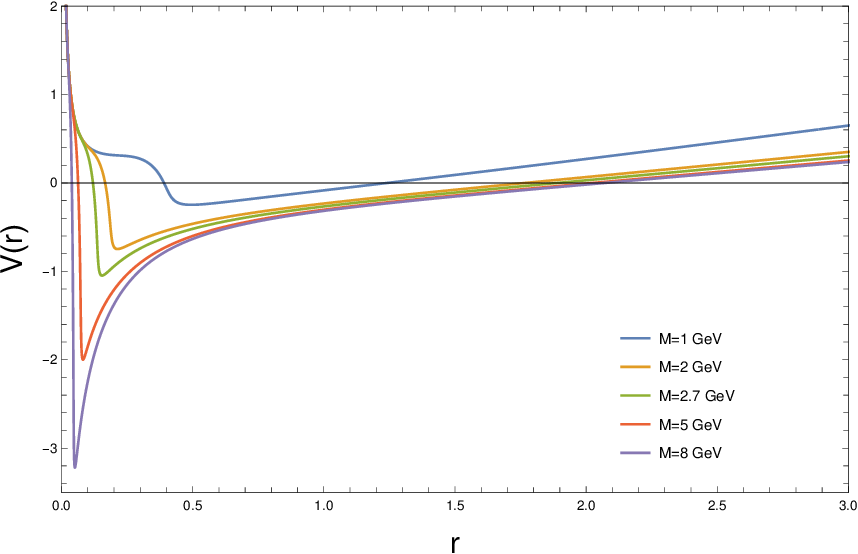}
    \caption{$q\bar{q}$ potential for different values of $M$.}
    \label{fig:fig5.1}
  \end{figure}
  \begin{figure}[h]
	\includegraphics[width=0.45 \textwidth]{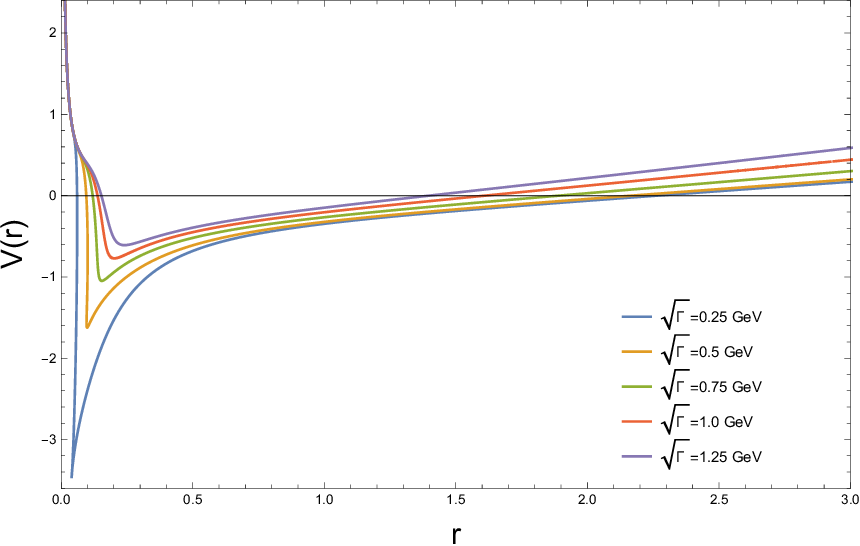}
    \caption{$q\bar{q}$ potential for different values of $\sqrt{\Gamma}$. }
    \label{fig:fig5.2}
  \end{figure}
We plot the potential as a function of the $q\bar{q}$ separation, in Figure  \ref{fig:fig5.1} we fix $\sqrt{\Gamma}=0,75GeV$ and take five representative values of $M$ while in Figure  \ref{fig:fig5.2} we fix $M=2,7GeV$ and take five representative values of $\sqrt{\Gamma}$. 
For very small $r$ we have a positive divergence that differs completely  from the Coulomb potential,  
at large distances we have the expected linear potential. This divergence in very short distances is a consequence of the $\tanh$ term in the dilaton and flip the slope of the potential in the small $r$ region. Such a divergence is expected in a rotating system of two particles, where it is associated with an repulsive centrifugal force that prevents the particles to colapse in the origin. In the present case the $q\bar{q}$ system is not rotating and the divergence is a consequence of the deformation of the AdS space in the small $z$ region by the $\tanh$ term in the dilaton.

\subsection{Finite temperature} 
\begin{figure}
	\includegraphics[width=0.45 \textwidth]{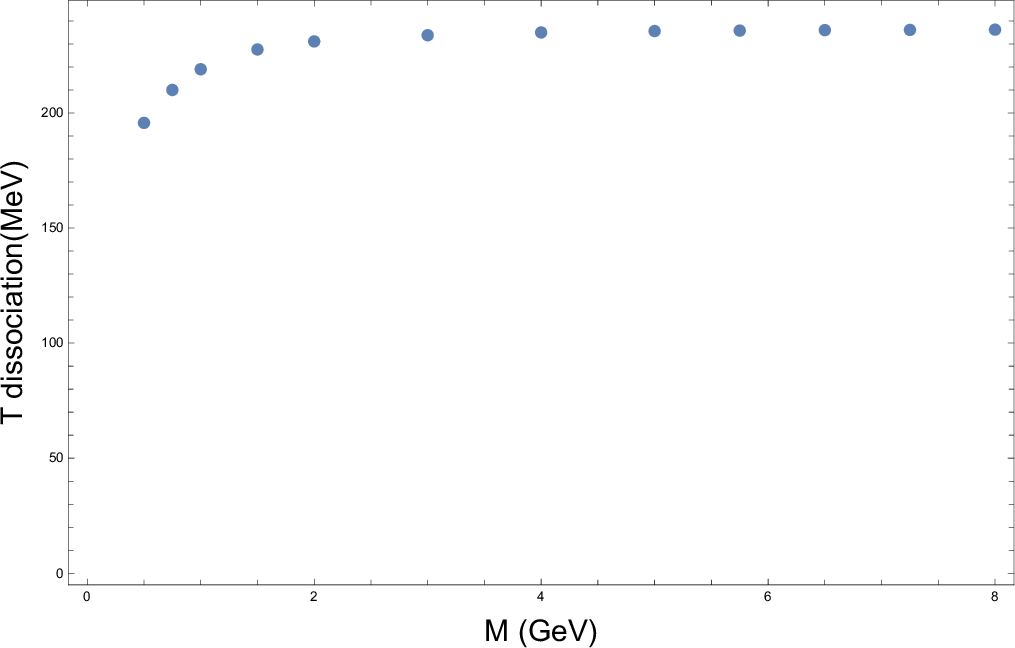}
    \caption{Dissociation temperature as a function of $M$.}
    \label{fig:fig6.1}
  \end{figure}

\begin{figure}
	\includegraphics[width=0.45 \textwidth]{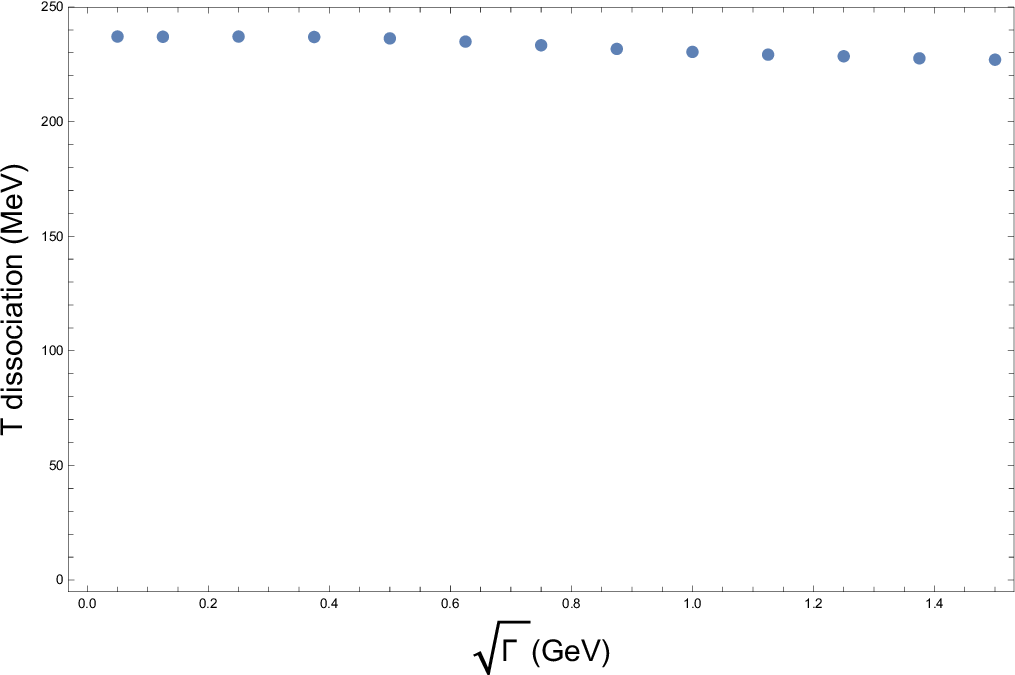}
    \caption{Dissociation temperature as a function of $\sqrt{\Gamma}$.}
    \label{fig:fig6.2}
  \end{figure}

We consider here the case of finite temperature in order to study the dissociation of the meson. 
For the present model the Euclidean black hole geometry is
\begin{equation}
 ds^2 = R^2\frac{e^{\Phi(z)}}{z^2}\left(f(z)dt^2 + d\vec{x}^2 + \frac{1}{f(z)}dz^2\right).
\end{equation}
The temperature of the dual gauge theory is   
\begin{equation}
 T  = \frac{1}{\pi z_h}.
\end{equation}
Taking into account the warp factor $f(z)$ present in the black hole geometry, we find
the $q\bar{q}$ separation distance as a function of  $z_m$:
\begin{equation}
r(z_m) = 2 z_m\int_0^1 dv \frac{v^2e^{[\Phi(z_m)-\Phi(v z_m)]}}{f(vz_m)\sqrt{1-v^4e^{2[\Phi(z_m)-\Phi(v z_m)]}\frac{f(z_m)}{f(vz_m)}}}.
\end{equation}
Looking at the function $r(z_m)$ while decreasing the location of the event horizon we determine the dissociation temperature for charmonium.
The  dissociation temperature is obtained for different values of the additional scales  introduced in the model, in  Figure \ref{fig:fig6.1}  we 
show the sensitivity of the dissociation temperature with $M$ and in Figure \ref{fig:fig6.2} we show the sensitivity with $\sqrt{\Gamma}$.

The dissociation temperature varies smoothly with the additional energy scales introduced in the model. 
The observed behavior of the critical temperature, with both $M$ and $\sqrt{\Gamma}$,
is directly related with the observed behavior of  the location of the infrared wall.
The dissociation is reached when the horizon gets close to the infrared wall. This connection is clear when comparing the results 
in Figure \ref{fig:fig4.1} and Figure \ref{fig:fig6.1}, as well as in Figure \ref{fig:fig4.2} and Figure \ref{fig:fig6.2}.
Even that the infrared wall does not determine completely
the location $z_{critical}$ where the dissociation occurs there is a monotonic relation between them.
We also face the situation that the  location of the event horizon, where
the dissociation occurs, is always bigger then the infrared wall location, meaning that the string breaks down before it touches the horizon.  
 
\subsection{The refined model}
\begin{figure}[h]
	\includegraphics[width=0.45 \textwidth]{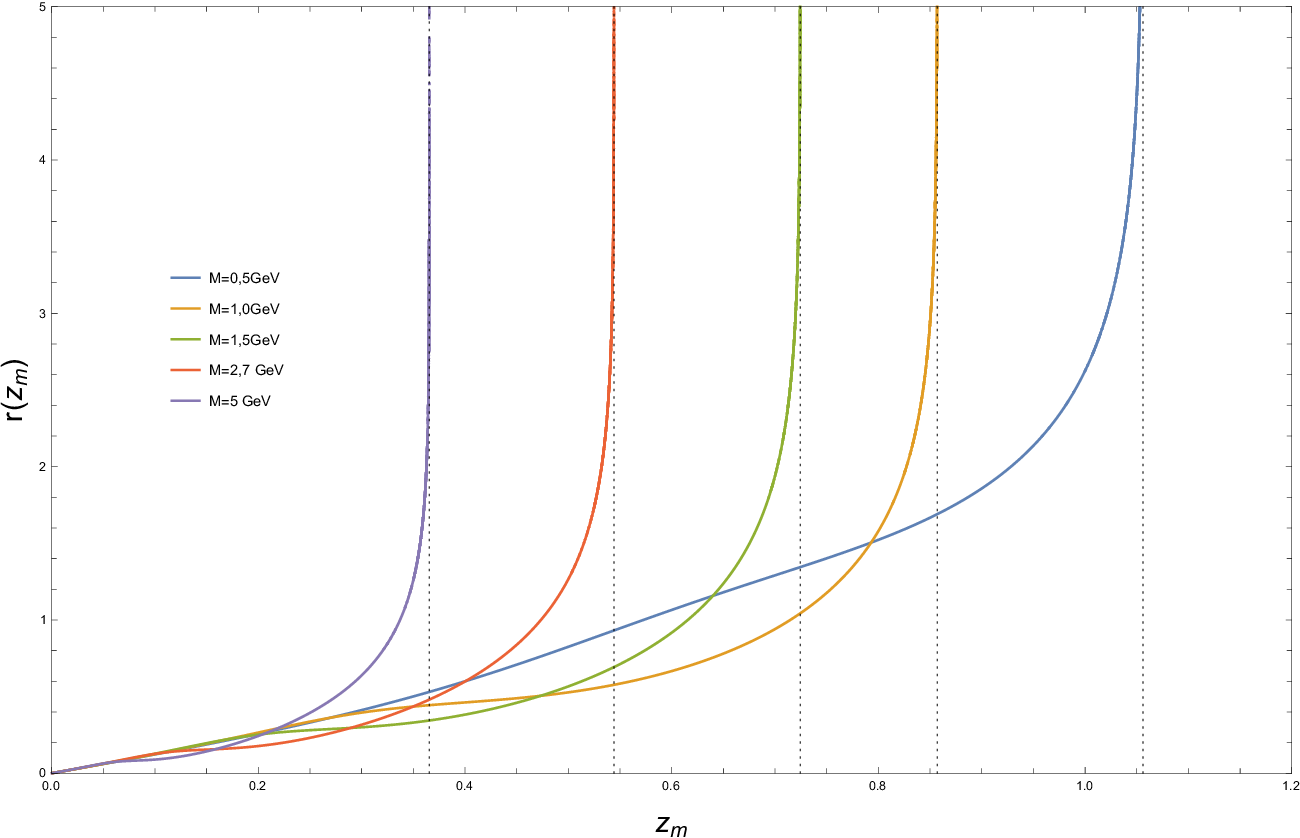}
    \caption{The infrared wall as a function of $M$ in the refined model.}
    \label{fig:fig7.1}
  \end{figure}

\begin{figure}[h] 
	\includegraphics[width=0.45 \textwidth]{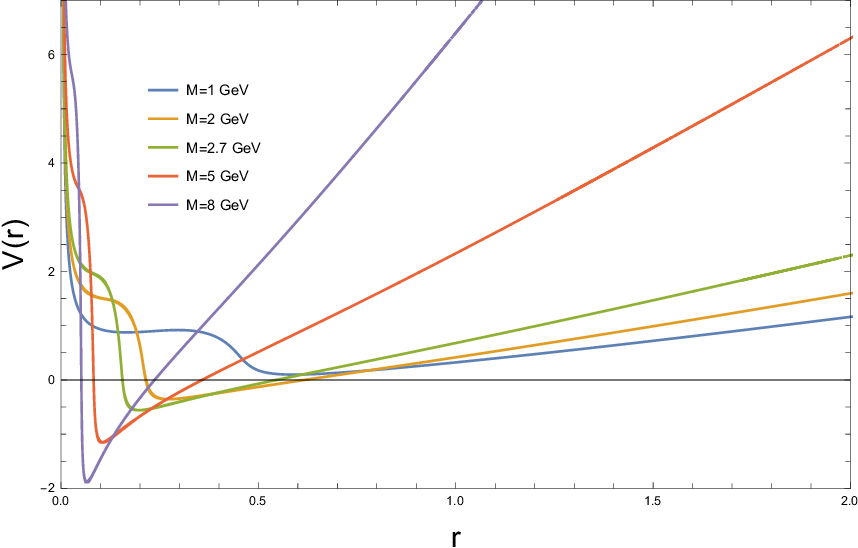}
    \caption{$q\bar{q}$ potential for different values of $M$ in the refined model.}
    \label{fig:fig7.2}
  \end{figure}

\begin{figure}[h] 
	\includegraphics[width=0.45 \textwidth]{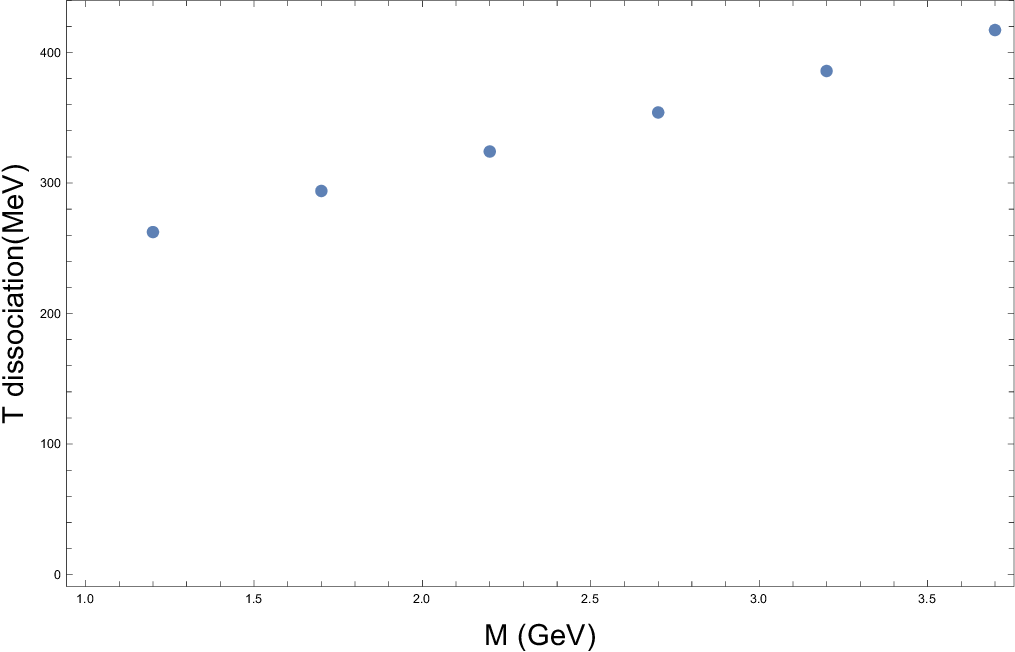}
    \caption{Dissociation temperature as a function of $M$ in the refined model.}
    \label{fig:fig7.3}
  \end{figure}

The AdS/QCD model of \cite{Braga:2017bml} is refined in Ref. \cite{Braga:2018zlu} to improve the description of heavy meson spectrum. The refinement consists in the introduction of a linear term in the dilaton profile, obtaining the dilaton 
\begin{equation}
\tilde{\Phi}(z) =  k^2z^2+ Mz + tanh \left( \frac{1}{Mz}-\frac{k}{\sqrt{\Gamma}} \right).  \label{refineddilaton}    
\end{equation}
The results obtained in this section are extended to this refined model replacing $\Phi\to \tilde{\Phi}$. The linear term in eq.(\ref{refineddilaton}) leads to a strong dependency of the world-sheet configurations on the energy scale fixed by $M$. We explore the sensitivity of the refined model on $M$ and show in Figure \ref{fig:fig7.1}   the location of the infrared wall, in Figure \ref{fig:fig7.2}  the $q\bar{q}$ potential and in Figure \ref{fig:fig7.3} the dissociation temperature of the meson. The vertical dotted lines in Figure \ref{fig:fig7.1} are placed in the location of the infrared wall, determined by the global minimum of $f_{\tilde{\Phi}}(z) = e^{2\tilde{\Phi}(z)}/z^4$.
  
In the present case the interaction potential is also divergent in origin, for large separation the potential is linear and in the present case the string tension grows monotonically with $M$.  The location of the infrared wall is very sensitivity on $M$, decreasing when $M$ increases. As a consequence, the dissociation temperature is also very sensitivity on $M$, increasing linearly. The addition of the linear term in eq.(\ref{refineddilaton}) makes $M$ an energy scale that affect simultaneously the ultraviolet and the infrared regions of the bulk geometry. The role of $\sqrt{\Gamma}$ is not affected by the presence of the linear term.

\section{Discussion}
 Decay constants are associated with the ultraviolet region in momentum space and recent results makes it clear that a non-trivial spectrum of decay constants requires a dilaton profile that does not vanishes at the boundary of the bulk geometry \cite{MartinContreras:2019kah}. In the soft case where there is no cutoff, as it happens for the models discussed in the previous section, to precise cancel the ultraviolet divergence one cannot ignore the dilaton there. Here an alternative regularization of the string action was proposed, this regularization cancel the ultraviolet divergence without generating an infrared one. 
 The regularization proposed here  reproduces the Cornell potential when applied to the soft wall model and differs from the one presented in Ref. \cite{Andreev:2006ct} by a small negative shift that is constant for $r\to \infty$ and is zero for $r\to 0$. 
For the models analyzed here the Cornell potential is not obtained, in the model with ultraviolet cutoff the potential is finite at $r=0$ and its first derivative is not defined there. The failure in the first derivative is a characteristic of the hard cutoff and also appears in the hard wall model, where the first derivative of the dual potential is not defined at the critical separation where the string touches the hard wall. For the models with modified dilaton profiles the potential is divergent for $r\to0$, this divergence at the origin is intriguing for a system that is not rotating.  A deeper understanding of these potentials is needed and will be subject for future work.  
 
 It is also obtained here that holographic calculations for static $q\bar{q}$ pairs lead to a dissociation temperature that is not so sensitive to the additional energy scales for the models of Refs. \cite{Evans:2006ea,Afonin:2011ff, Braga:2015jca} and \cite{Braga:2017bml}. These models implement deformations in the quadratic dilaton background that affect mostly the ultraviolet region but almost does not deform the infrared region. In the model of Ref. \cite{Braga:2018zlu} the additional energy scale $M$ deforms simultaneously the infrared and the ultraviolet regions, and in this model the dissociation temperature of the dual $q\bar{q}$ pair is very sensitive on this additional scale. This sensitivity is expected from the spectral functions calculated using these AdS/QCD backgrounds and the present results suggests that  the model of Ref. \cite{Braga:2018zlu}  is more consistent in describing heavy mesons. 
 Even that the present analysis favor the model of Ref. \cite{Braga:2018zlu} over the models of Refs. \cite{Evans:2006ea,Afonin:2011ff, Braga:2015jca} and \cite{Braga:2017bml}, it presents a positive divergence of the static potential at small distances not reproducing the expected Cornell potential.  This divergence changes the slope of the $q\bar{q}$ potential for small separation, violating the condition that it is everywhere  non-decreasing \cite{Bachas:1985xs} and disfavoring the corresponding holographic backgrounds as a model for heavy mesons. 
 A natural question is that if this unexpected divergence is a peculiar fact of the analysed models or it will be present in any AdS/QCD model that describes the observed decay constants of the  heavy vector mesons.

\acknowledgments
 The author  thanks to M. A. Martin Contreras for nice  discussions and important suggestions.  The author also thanks to Eduardo Folco Capossoli and Henrique Boschi-Filho for important communications.

\end{document}